\documentstyle[12pt,epsf]{article}
\def\Journal#1#2#3#4{{#1} {\bf #2}, #3 (#4)}
\def\RNC{\em Riv. Nuovo Cimento}
\def\NCA{\em Nuovo Cimento}
\def\NPB{{\em Nucl. Phys.} B}
\def\PTP{{\em Prog. Theor. Phys.}}

\def\PRL{\em Phys. Rev. Lett.}
\def\PR{{\em Phys. Rev.}}

\def\PRD{{\em Phys. Rev.} D}

\def\RMP{{\em Rev. Mod. Phys.}}

\def\PREPC{{\em Phys. Rep.} C}
\def\RNC{{\em Riv. Nuovo Cimento}}

\newcommand{\be}{\begin{equation}}
\newcommand{\ee}{\end{equation}}
\newcommand{\bea}{\begin{eqnarray}}
\newcommand{\eea}{\end{eqnarray}}

\newcommand{\nonu}{\nonumber\\}

\begin{document}
\hyphenation{com-me-nsu-rate-in-com-me-nsu-rate}
\unitlength1cm
\begin{center}
{\LARGE RG generated fermion mass}\\
\vspace*{1cm}
Vincenzo Branchina\footnote{branchin@physx22.u-strasbg.fr}\\
\vspace*{1cm}
Laboratory of Theoretical Physics, Louis Pasteur University\\
3 rue de l'Universit\'e 67087 Strasbourg, Cedex, France\\
\vspace*{1.5cm}
{\LARGE Abstract}\\
\end{center}

Chiral symmetry breaking in a purely fermionic theory is investigated by the 
help of the renormalization group method. The RG equation for 
the running mass $m_k$ admits a solution with vanishing bare mass and finite 
physical mass. The running Fermi coupling constant, $G_k$, converges to a 
finite (renormalized) physical value. It is also shown that the RG equation 
for $\tilde G$, the dimensionless Fermi coupling, has an UV fixed point 
$\tilde G_{UV}$. Contrary to a previous result however, it is proven that 
the chiral symmetry breaking point $\tilde G_c$ does not coincide with 
$\tilde G_{UV}$.

\vspace*{1cm}

The problem of the dynamical generation of a fermion mass has been studied 
over the years by several authors\cite{john}. Nambu and Jona-Lasinio (NJL) 
considered a purely fermionic model in the mean field approximation and 
found a non trivial solution for the fermion mass, for values of the Fermi 
constant larger than a critical one \cite{njl}.
The non perturbative nature of this approximation is obvious as the 
chiral symmetry of the model forbids the appearance of a mass term 
at any order of perturbation theory.

Here I study a fermionic model by the help of a different non-perturbative 
technique, the renormalization group method. It is widely recognized that 
the approach pioneered by Wilson\cite{wilsrg} provides a powerful method 
to study quantum and statistical field theories. 
In the Wegner-Houghton\cite{weg} realization it is implemented by 
establishing an exact integro-differential flow equation for the Wilsonian 
effective action $S_k$. 
If $\Lambda$ is an UV scale, in the $k\to\Lambda$ limit (and eventually 
$\Lambda\to\infty$) $S_k$ corresponds to the bare UV action, while $S_{k=0}$ 
is the physical effective action. 
The Wegner-Houghton equation as it stands is actually  
an intractable one, a systematic approximation scheme can be built 
by the help of the derivative expansion. At its lowest order 
it gives the so called Local Potential Approximation (LPA).  

By considering the LPA for a fermionic model with discrete $\gamma_5$ 
symmetry of the bare action and potential truncated to 
the quadrifermionic interaction term, I show 
(for the first time at the best of my knowledge) that : 

- The RG equation for the running mass, $m_k$, admits a solution 
vanishing in the $k\to\infty$ limit and finite at $k=0$, i.e. a 
solution breaking the discrete $\gamma_5$ symmetry of the bare theory ;

- The RG equation for the running Fermi coupling constant, $G_k$, admits a 
solution having the canonical scaling in the UV region, 
$G_k \sim \frac{1}{k^2}$, and flowing to a {\it finite} value in the IR, 
i.e. the theory can be renormalized.  

By freezing the dimensionless Fermi coupling constant to a given 
fixed value, I also consider the RG equation for the running mass alone.
According to the value of the Fermi constant, I find that  
a non vanishing physical mass, $m_{ph}=m_{k=0}\neq 0$, 
may or may not be generated from the chirally invariant bare theory. 
Clearly with this additional approximation we get close 
to the approach of the original NJL paper. There are however important  
differences between our approach and that of Ref.\cite{njl}. 

In this latter paper the quantum fluctuations (responsible for the dynamical 
generation of the fermion mass) are taken into account by the help of a mean 
field approximation that allows to establish the gap equation for a constant 
(momentum independent) mass function. As they were dealing with a 
(perturbatively) non renormalizable theory, the authors of Ref.\cite{njl} 
restricted themselves to consider the {\it cut-off} theory. 
Being the mass function approximated by a constant, the one-loop 
integral obviously receives a quadratically divergent contribution. 

In our approach the quantum fluctuations are taken into account 
through the resummation operated by the RG equation, the momentum dependence 
of the mass function being given by the running scale. The solution to our 
RG equation, $m_k$, vanishes in the 
UV and the $k\to \infty$ limit does not generate divergent contributions.
A quantitative and detailed comparison between our approach and the 
NJL approximation will be presented in a forthcoming paper\cite{me}. 
It is sufficient to add here that, within a certain approximation, 
our RG equation for the running mass $m_k$ reproduces the NJL result. 

A related aspect of our analysis is that of the existence of a continuum 
limit. 

In massless QED the lowest order approximation to the gap equation, 
for values of the fine structure constant $\alpha$ 
larger than a critical one, $\alpha_c=\frac{\pi}{3}$, gives 
for the dynamically generated fermion mass \cite{mira83} \cite{bardeen} : 

\be\label{mir}
m=\Lambda f(\alpha), 
\ee

\noindent
where $\Lambda$ is the UV cut-off and $f(\alpha)$ is a known function 
of $\alpha$. Miransky suggested\cite{mira85} that this equation should 
be regarded as defining the UV flow of the running coupling constant 
$\alpha=\alpha(\Lambda)$, the fermion mass $m$ being fixed. 
From the specific form of the function $f(\alpha)$ it is immediately 
found that $\alpha(\Lambda)$ flows toward $\alpha_c$ for $\Lambda\to\infty$.
This interpretation has received lot of attention as it seemed the 
only possibility to evade the conclusion that the physics of 
the dynamical generation of a fermion mass actually occurs at the 
cut-off scale\cite{bardeen2}. Miransky also extended it 
to other gauge theories as well as to the quadrifermionic theory \cite{mira}. 
Following this suggestion several models have been constructed and the phase 
structure of certain theories investigated.

In the quadrifermionic theory the result of the mean field approximation
can be written again as $m=\Lambda f(\tilde G)$, where $\Lambda$ is the UV 
cut-off, $\tilde G$ is the dimensionless Fermi coupling and $f(\tilde G)$ 
a known function of $\tilde G$. The UV flow of the Fermi constant 
$\tilde G=\tilde G(\Lambda)$ is defined as before and 
accordingly it is found that the critical point $\tilde G_c$ coincides with 
an UV fixed point $\tilde G_{UV}$ \cite{mira}. 

This interpretation however is not the result of a RG analysis. It rather 
comes from the attempt to define a continuum limit out of 
the results of the gap equation. 

By studying for the first time this problem in a real RG framework, I  
show that contrary to the Miransky suggestion the two points $\tilde G_c$ 
and $\tilde G_{UV}$ are well separated, $\tilde G_c$ being smaller 
than $\tilde G_{UV}$.  
 
The (euclidean) wilsonian action for our model at the scale $k$ in 
the LPA  has the form :

\be\label{lpa}
S_k=\int d^4x \Bigl[ {\overline\psi}\gamma_\mu\partial_\mu\psi + 
U_k({\overline\psi}\psi)\Bigr].
\ee

\noindent
Expanding  the potential $U_k({\overline\psi}\psi)$ in powers of 
$\bar\psi\psi$ and retaining terms up to $(\bar\psi\psi)^2$, i.e. up to 
the quadrifermionic interaction term,  

\be\label{expa}
U_k(\bar\psi\psi) = m_k \bar\psi\psi - \frac{G_k}{2} (\bar\psi\psi)^2,
\ee

\noindent
the RG equation for $S_k$ actually reduces to a system of differential 
equations for the running mass $m_k$ and the running Fermi coupling constant 
$G_k$. 

The goal is to see whether a solution for $m_k$ exists such that for a 
vanishing bare mass $m_B$, the UV limit of the running mass, a finite 
physical mass, $m_{ph}=m_{k=0}=finite$, is obtained.

It is often convenient to move to dimensionless variables. 
Introducing the dimensionless scale parameter $t=ln\frac{\mu}{k}$, 
where $\mu$ is a given boundary  value of $k$, together with the 
dimensionless running mass $\tilde m_t$ and running Fermi coupling constant 
$\tilde G_t$, 

\be\label{dimles}
\tilde m_t=\frac{m_k}{k} ~~~~~~~~~~{\rm and}~~~~~~~~~~
\tilde G_t=k^2 G_k,
\ee

\noindent
the flow equations for $\tilde m_t$ and $\tilde G_t$, 
obtained after integration of the degrees of freedom in the momentum shell 
$[t,t+\delta t]$ and taking the $\delta t \to 0$ limit, are :

\bea
\frac{d \tilde m_t}{d t}&=&\tilde m_t\Biggl[1+\frac{3}{8\pi^2}
\frac{\tilde G_t}{(1+m^2_t)} \Biggr]\label{eqm}\\
\frac{d \tilde G_t}{d t}&=&-2 \tilde G_t\Biggl[1 - \frac{1}{8\pi^2} 
\frac{\tilde G_t}{(1+m^2_t)^2} \Biggr].\label{eqg}
\eea

At a first sight it could seem from Eq.({\ref{eqm}) that given the boundary 
$m_{t_0}=0$ at an UV scale $t_0$, the only possible solution 
to this equation is $\tilde m_t=0$ for any value of $t$. This in turn would 
imply that $m_{ph}=0$, the symmetry is not broken and no mass term is 
generated.  

\noindent
The above reasoning is correct if this boundary value
is assigned at a finite value $t_0$. If however it is given for 
$t_0 \to -\infty$ (that corresponds to the UV limit $k \to \infty$), 
a symmetry breaking solution $m_{ph} = finite$ exists.

The complete analysis of the system (\ref{eqm})-(\ref{eqg}) can be 
performed only numerically. However the UV ($t\to -\infty$) and the 
IR ($t\to +\infty$) asymptotic regions can be studied analytically. 

Let's start with the UV region and 
suppose that a solution exists such that for $t\to -\infty$, 
$m^2_t$ vanishes more rapidly than $\tilde G_t$. Under this assumption the 
system becomes :

\bea\label{sy}
\frac{d \tilde m_t}{d t}&=&\tilde m_t\Biggl[1+\frac{3 \tilde G_t}{8\pi^2}
\Biggr]\label{eqmas}\\
\frac{d \tilde G_t}{d t}&=&-2 \tilde G_t\Biggl[1 - \frac{\tilde G_t}{8\pi^2} 
\Biggr]\label{eqgas}
\eea

\noindent
Eq.(\ref{eqgas}) has the solution : 

\be\label{solg}
\tilde G_t=\frac{8\pi^2}{1 + C e^{2t}},
\ee

\noindent 
where $C$ is an integration constant. We note here that, in terms of the
dimensionful Fermi constant, this UV flow is :

\be\label{dimg}
G_k \sim_{k\to\infty}\frac{8\pi^2}{k^2}.
\ee

\noindent
To study the behavior of Eq.(\ref{eqmas}) in the $t\to -\infty$ region, 
we can now replace in this equation the asymptotic value 
$ \tilde G_t\sim 8\pi^2$. Eq.(\ref{sy}) then  becomes :

\be\label{eqmasy}
\frac{d \tilde m_t}{d t} - 4 \tilde m_t = 0
\ee

\noindent
whose solution is trivially (A is an integration constant) :  

\be\label{solm}
\tilde m_t = A e^{4t}.
\ee

\noindent
The assumption that lead us to the approximated system 
(\ref{eqmas})-(\ref{eqgas}) is consistent with Eqs.(\ref{solg}) and 
(\ref{solm}), consequently these are asymptotic solutions to the original 
system (\ref{eqm})-(\ref{eqg}). We also see that the system 
(\ref{eqm})-(\ref{eqg}) possesses the UV ($t=-\infty$) fixed point :

\be\label{fp} 
{\tilde m}_{UV}=0 ~~~~~~~~~~,~~~~~~~~~~{\tilde G}_{UV}=8\pi^2.
\ee

\noindent
Moreover, as $k=\mu e^{-t}$, from Eq.(\ref{solm}) we have : 

\be\label{mas}
m_k \sim_{k\to\infty} ~A\frac{\mu^4}{k^3},
\ee

\noindent
that gives :  $m_B=lim_{k\to\infty} m_k = 0$.  

\noindent
Eq.(\ref{mas}) is a potentially interesting result, we have found
a running mass $m_k$ vanishing in the $k\to \infty$ limit.

Let's move now to the IR region.
By  simple inspection we see that in the asymptotic region $t \to \infty$, 
a solution to the system (\ref{eqm})-(\ref{eqg}) exists such that :

\bea
 \tilde m_t &\sim_{t \to \infty} &e^t\label{eq1}\\
 \tilde G_t &\sim_{t \to \infty} & e^{-2t}\label{eq2}.
\eea

\noindent
This again is a potentially interesting result. Moving to
dimensionful parameters, Eqs.(\ref{eq1}) and (\ref{eq2}) in fact give :

\bea
m_{ph} &=\lim_{k \to 0} m_k = finite \label{equ1}\\
G_{ph} &=\lim_{k \to 0} G_k = finite \label{equ2}.
\eea

\noindent
We don't know yet however whether a solution $\{\tilde m_t,\tilde G_t\}$ to 
the system  (\ref{eqm})-(\ref{eqg}) exists possessing both the IR and the 
UV behavior respectively given by Eqs.(\ref{eq1})-(\ref{eq2})
and (\ref{solg})-(\ref{solm}). 
That such a solution exists is the main result of this paper and will be 
proven numerically later.

Before moving to the numerical solution of the system (\ref{eqm})-(\ref{eqg})
however, we want to consider an additional approximation under which it 
is possible to find an analytical solution. We freeze the value of the 
Fermi constant to a given fixed value and restrict ourselves to consider 
the evolution of the running mass $\tilde m_t$ alone. 

This approximation to our RG equations is 
already worth to study. In the RG framework, the flow equation for the 
running mass plays the same role as the gap equation for the mass function
in the Schwinger-Dyson approach. Moreover, as we have pointed out before,
it can be proven \cite{me} that in a certain limit it 
reproduces the mean field result.

Let's freeze $\tilde G_t$ to its UV fixed point value found above, 
namely $\tilde G_t={\tilde G_{UV}}=8\pi^2$. 
Under this approximation our original system (\ref{eqm})-(\ref{eqg}) 
reduces to the differential equation for $\tilde m_t$ :

\be\label{meqt}
\frac{d \tilde m_t}{d t} = \tilde m_t\Bigl[1+\frac{3}{1+\tilde m_t^2}\Bigr],
\ee

\noindent
that we can conveniently rewrite as,

\be\label{m2}
\frac{d \tilde m_t^2}{d t} =
2 \tilde m_t^2\Bigl[1+\frac{3}{1+\tilde m_t^2}\Bigr].
\ee

We have already seen that a solution to Eq.(\ref{m2}) exists 
such that in the UV limit, i.e. for $t\to - \infty$, 

\be\label {eqm2} 
\tilde m_t^2 \sim e^{8t}
\ee

\noindent
(see Eq. (\ref{solm}) above). 

We can also immediately verify that in 
the IR, i.e. for $t\to \infty$, Eq.(\ref{m2}) has the asymptotic solution 
(B is an integration constant)

\be\label{iras}
\tilde m_t^2 = -3 + B e^{2t} \sim B e^{2t}.
\ee

\noindent
Moving to the dimensionful running mass we have : 

\be\label{mph}
m_{ph}^2= lim_{k\to 0} m_k^2 = B\mu^2=finite. 
\ee

\noindent
Once more, we don't know yet whether a solution to Eq.(\ref{m2})
exists with both the UV and the IR behavior of Eqs.(\ref{eqm2}) and 
(\ref{iras}) respectively. Fortunately the analytical solutions to 
Eq.(\ref{m2}) can be found and one of them is relevant to our problem. 
It has a quite long expression that we can write as :

\be\label{man}
\tilde m_t^2=\frac{\sqrt 3}{6}
\Biggl[\frac{b_t^{\frac{1}{2}}}{a_t^{\frac{1}{6}}}
 +\frac{\Bigl[24 a_t^{\frac{1}{3}} b_t^{\frac{1}{2}}
 -3 a_t^{\frac{2}{3}} b_t^{\frac{1}{2}}
 +4 e^{8(t+\alpha)} b_t^{\frac{1}{2}} + 32 c_t\Bigr]^{\frac{1}{2}}}  
 {a_t^{\frac{1}{6}} b_t^{\frac{1}{4}}}\Biggr] -3
\ee

\noindent
where 

\bea
a_t&=&8 e^{8(t+\alpha)}\Bigl[\frac{\sqrt 3}{9}
\biggl( e^{8(t+\alpha)} +27 \biggr)^{\frac{1}{2}} - 1 \Bigr]\nonu
b_t&=&12 a_t^{\frac{1}{3}} + 3 a_t^{\frac{2}{3}} - 4 e^{8(t+\alpha)}\nonu
c_t&=&\sqrt 6 e^{4(t+\alpha)}    
\sqrt{\sqrt 3 ( e^{8(t+\alpha)} +27 )^{\frac{1}{2}} - 9 },\label{abc}
\eea

\noindent
and $\alpha$ is an integration constant.

A long but straightforward computation shows that the above solution, 
Eq.(\ref{man}), has the required IR and UV asymptotic behavior. 

This is one of our main results. 
The RG equation for the running mass $m_k$ admits a solution such 
that $lim_{k\to\infty} m_{k} = 0$ and $lim_{k\to 0} m_{k} = finite$. 
In other words {\it{the RG equation for $m_k$ generates a physical mass from 
the massless bare theory, thus breaking dynamically the chiral symmetry}}. 
Through a cross-over region the UV $\frac{1}{k^3}$ flow of the mass function  
is converted into an IR scaling giving rise to a finite value at $k=0$. 
 
Of course the question arises concerning the existence of a critical 
value for the Fermi coupling constant. 

We have just seen that for $\tilde G=8\pi^2$ we have a symmetry breaking 
solution. As compared to the self consistent 
approach\cite{njl} we do not have here an algebraic equation for 
$m_{ph}$ but rather a differential equation for $\tilde m_t$ 
(or, what amounts to the same thing, for $m_k$). We expect that 
there exists a critical value $\tilde G_c$ of the dimensionless Fermi 
constant $\tilde G$ such that for  $\tilde G < \tilde G_c$ we have 
$m_{ph} = 0$, while for $\tilde G > \tilde G_c$ it is $m_{ph} \neq 0$.
To find  $\tilde G_c$ we should in principle replace $\tilde G = 8\pi^2$ 
in Eq.(\ref{m2}) with a generic value of $\tilde G$ and seek for the 
solution $\tilde m_t$. Unfortunately it is not a trivial task 
to look for analytical solutions of Eq.(\ref{m2}) for arbitrary values 
of $\tilde G$. Analytical solutions can only be obtained for certain 
specific values of $\tilde G$.  

It is now not too difficult to check that for $\tilde G = 4\pi^2$ and 
$\tilde G = \frac{8\pi^2}{3}$ a solution $\tilde m_t$ vanishing in the 
UV and converging to a finite value in the IR exists, while for the 
value $\tilde G = 2\pi^2$ this solution is lost. Moreover, while for 
$\tilde G = 8\pi^2$ we have found that the UV behavior of the mass function 
is $m_k \sim \frac{1}{k^3}$, for  $\tilde G=4\pi^2$ it is 
$m_k \sim \frac{1}{k^2}$ and for $\tilde G=\frac{8\pi^2}{3}$ it is 
$m_k \sim \frac{1}{k}$. We are then lead to the conclusion that 
$\tilde G_c$ lies in the range $[2\pi^2,\frac{8\pi^2}{3}]$ 
\footnote{From the specific form of the solution at 
$\tilde G=\frac{8\pi^2}{3}$, I believe that $\tilde G_c=\frac{8\pi^2}{3}$.
In any case the precise location of the critical point is not our main 
concern here. What is important is the result that the UV fixed point, 
$\tilde G_{UV}$, and the critical point, $\tilde G_c$, do not 
coincide.}. 

This is another important result. Comparing with our previous result 
$\tilde G_{UV} = 8 \pi^2$, we conclude that the UV fixed 
point $\tilde G_{UV}$ does not coincide with the critical point $\tilde G_c$.

Needless to say the existence and the location of an UV fixed point for 
$\tilde G_t$ can only be established from the RG equation for $\tilde G_t$ 
itself. Indications coming from other arguments have to be taken with 
caution. Our RG analysis has shown that the theory possesses 
an UV fixed point, $\tilde G_{UV}$, that however does not coincide with the 
Miransky limit \cite{mira}. In addition the correct UV scale dependence of 
$\tilde G_t$, Eq.(\ref{solg}), is different from the one deduced from this 
interpretation of the gap equation. As the 
above criticism obviously extends to any other theory where this interpretation
was applied, we claim that those phase diagrams and models based on it have to 
be reconsidered. 

\begin{figure}
\begin{minipage}{6cm}
	\epsfxsize=4.5cm
	\epsfysize=6cm
 	\centerline{\epsffile{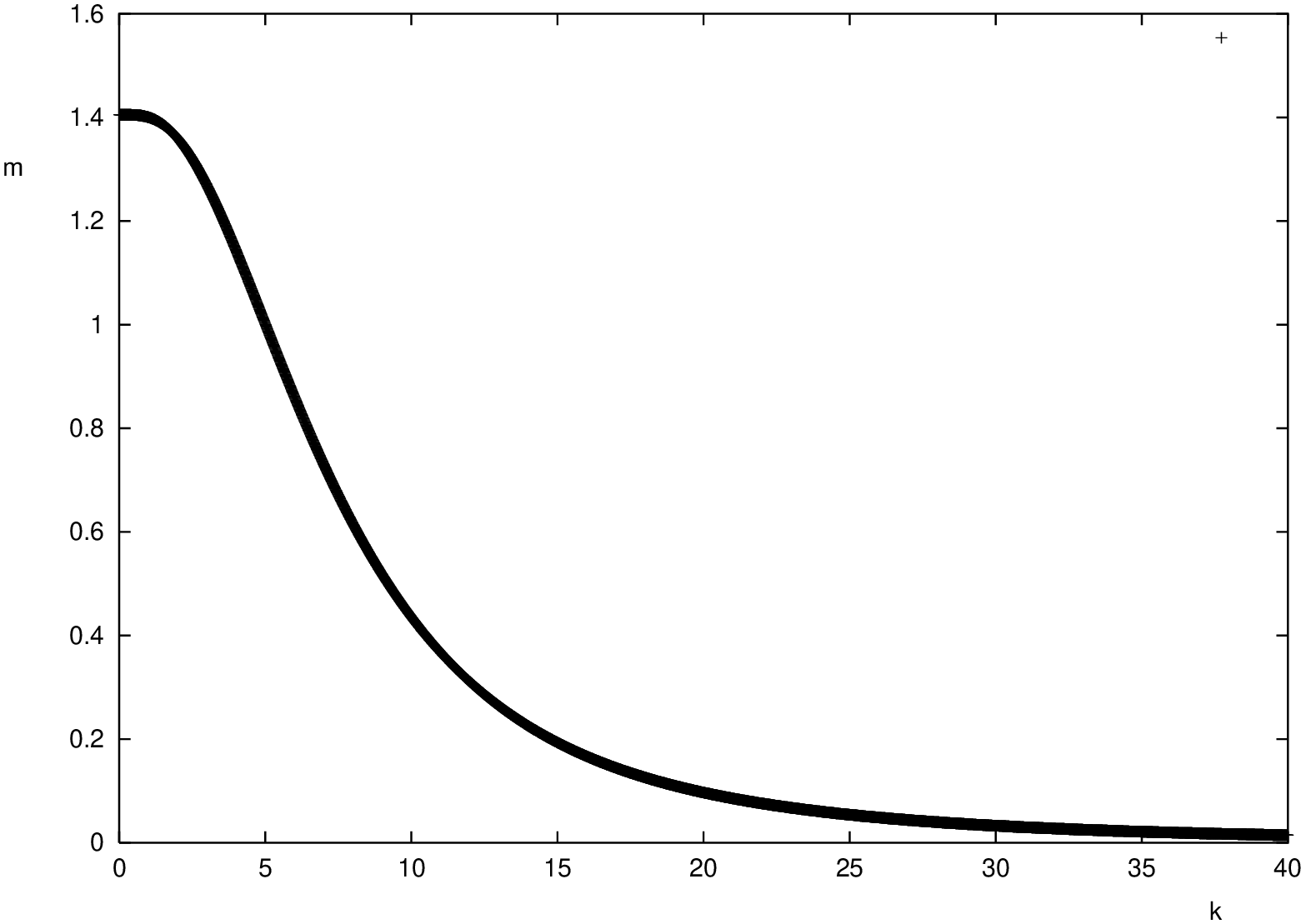}}
\centerline{(a)}
\end{minipage}
\hfill
\begin{minipage}{6cm}
	\epsfxsize=4.5cm
	\epsfysize=6cm
	\centerline{\epsffile{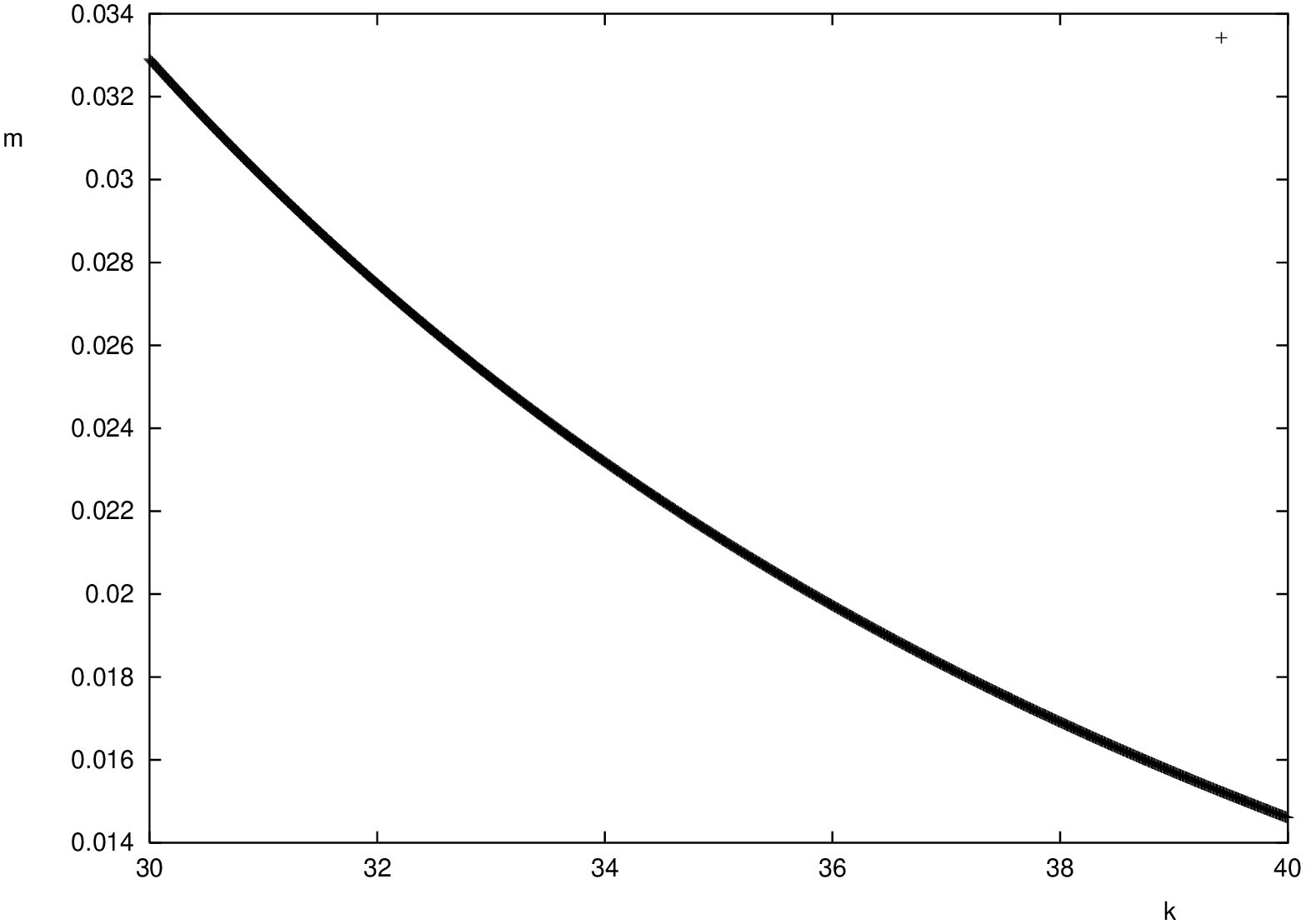}}
\centerline{(b)}
\end{minipage}
\caption{(a): In this figure we see the running of the dimensionful mass 
$m_k$, the boundary condition is given in the text. (b): Here we focus 
on the flow well on the right of the crossover region where we recognize 
the UV scaling, $m_k \sim \frac{1}{k^3}$.}
\end{figure}

We conclude now by presenting the numerical solution of the 
system (\ref{eqm})-(\ref{eqg}). 
As it is preferable to work directly with the 
dimensionful parameters, in Figs.(1) and (2) we show the running 
of $m_k$ and $G_k$ versus $k$. As boundary values we have taken 
$m_k=10^{-6}$ and $G_k=7.895 10^{-5}$ at $k=10^3$.

In Fig.(1.a) we see the running of the mass parameter $m_k$ and observe the 
transition from the UV regime where $m_k\to 0$ to the IR regime where 
$m_k$ converges to a finite value, the physical mass $m_{ph}$. 
The UV and IR asymptotic flows both coincide with our 
previous analytical results, Eqs.(\ref{mas}) and (\ref{iras}) respectively. 
Fig.(1.b) presents a magnification of the UV region where the 
$1/k^3$ UV behavior, found analytically in Eq.(\ref{mas}), is easily 
recognized. 

Fig.(2.a) shows the running of $G_k$ versus $k$. In the UV region its flow 
is nothing but the UV $1/k^2$ canonical scaling, already found in 
Eq.(\ref{dimg}). 
Through a crossover region this flow is converted to an IR scaling 
and $G_k$ converges to a finite value at $k=0$, as seen in Eq. (\ref{equ2}). 
As for $m_k$ we have 
magnified the UV region to better show the UV $1/k^2$ scaling.

\begin{figure}
\begin{minipage}{6cm}
	\epsfxsize=4.5cm
	\epsfysize=6cm
	\centerline{\epsffile{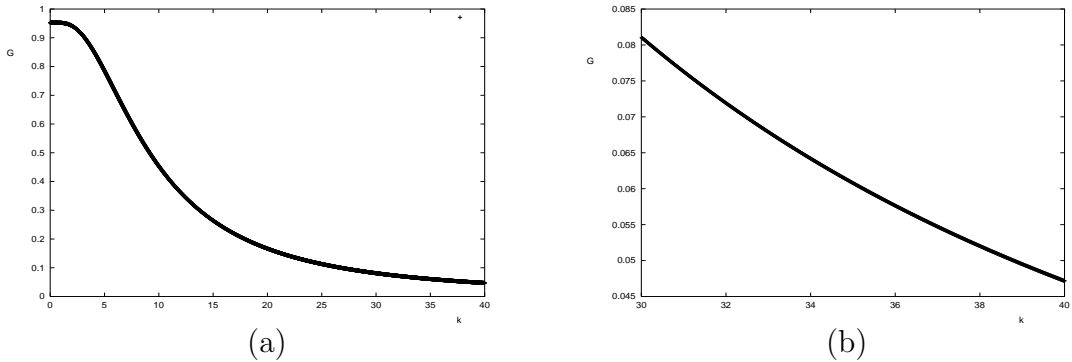}}
\centerline{(a)}
\end{minipage}
\hfill
\begin{minipage}{6cm}
	\epsfxsize=4.5cm
	\epsfysize=6cm
	\centerline{\epsffile{ag30.eps}}
\centerline{(b)}
\end{minipage}
\caption{ (a): In this figure we show the running of the dimensionful Fermi 
coupling constant $G_k$. (b): We focus here on the flow on the right of the 
crossover region and observe the UV scaling, $G_k \sim \frac{1}{k^2}$.} 
\end{figure}

In summary we have found that the RG equation for the running mass $m_k$
admits a solution that breaks the original discrete $\gamma_5$ chiral
symmetry of the bare theory and that the running Fermi coupling constant  
$G_k$ has the canonical scaling, $G_k \sim \frac{8\pi^2}{k^2}$, in the UV and 
flows to a renormalized value, $G_{k=0} = finite$, in the IR. 
It should not be underestimated here that for a theory in $d=4$ dimensions
we have established non-perturbative renormalization group equations allowing 
to follow the renormalization flow of the running coupling constants all the 
way down from the UV to the IR. The theory, at least within the approximation
considered in this paper, can be renormalized. There is certainly no need to 
remind the kind of (UV or IR) pathologies that are typically encountered 
in RG equations, think for instance of QED, $\lambda\phi^4$ or QCD. 

Finally we have also shown that an old result, concerning the
coincidence between the chiral symmetry breaking point and the UV fixed 
point of the theory, actually turns out to be incorrect.

\vspace*{0.3cm}
I would like to thank Veronique Bernard for many useful discussions.

\vfill
\eject

\end{document}